\documentclass{article}

\usepackage{arxiv}

\usepackage[utf8]{inputenc} 
\usepackage[T1]{fontenc}    
\usepackage{hyperref}       
\usepackage{url}            
\usepackage{booktabs}       
\usepackage{amsfonts}       
\usepackage{nicefrac}       
\usepackage{microtype}      
\usepackage{lipsum}
\usepackage{graphicx}
\usepackage{siunitx}
\usepackage{subfig}
\usepackage{caption}
\usepackage{amsmath}

\title{A novel accelerating structure based on a tapered parallel-plate waveguide with an integrated dielectric terahertz-driven accelerator}

\author{
 Andr\'es Leiva Genre \\
  Institute of Physics\\
  University of P\'ecs\\
   7624 P\'ecs, Hungary\\
  \texttt{andreslg@fizika.ttk.pte.hu} \\
   \And
 Zolt\'an Tibai \\
 Institute of Physics\\
  University of P\'ecs\\
  7624 P\'ecs, Hungary\\
  \And
  Luis Nasi\\
 Institute of Physics\\
 Szent\'agothai Research Center\\
  University of P\'ecs\\
  7624 P\'ecs, Hungary\\
  \And
  M\'aty\'as Kiss\\
  Institute of Physics\\
  University of P\'ecs\\
  7624 P\'ecs, Hungary\\
  \And
  G\'abor Alm\'asi\\
 Institute of Physics\\
 Szent\'agothai Research Center\\
  University of P\'ecs\\
  7624 P\'ecs, Hungary\\
  \And
  J\'anos Hebling\\
 Institute of Physics\\
 Szent\'agothai Research Center\\
  University of P\'ecs\\
  7624 P\'ecs, Hungary\\
  \And
  Szabolcs Turn\'ar\\
 Institute of Physics\\
 Szent\'agothai Research Center\\
  University of P\'ecs\\
  7624 P\'ecs, Hungary\\
  \\
}

\begin{document}
\maketitle
\begin{abstract}
We present a novel dielectric terahertz-driven accelerator (DTA) that integrates a dual-pillar grating structure within a tapered parallel-plate waveguide (TPPWG). This compact setup enables efficient particle acceleration using multi-cycle, narrowband terahertz (THz) pulses. The TPPWG serves a dual role: it enhances the THz field via geometric tapering and delivers it to the dielectric structure by efficient coupling. Experimental validation of the THz field inside the waveguide is conducted using electro-optic sampling. Optimization of waveguide parameters through time-domain simulations reveals a sixfold peak electric field amplification at the end of the waveguide. The dielectric accelerator is tailored for maximum acceleration by adjusting the DTA pillar radius and vacuum channel gap for relativistic electron beams. Particle-in-cell (PIC) simulations demonstrate that the structure supports net acceleration with gradients up to \SI{120}{\MeV \per \m} for \SI{0.1}{\giga \volt \per \m} field strengths, and can accommodate bunch charges up to 10 pC with minimal degradation. Energy spread evolution and beam dynamics are discussed in detail, including the role of phase slippage and bunch length. This work establishes the DTA-integrated TPPWG as a compact and scalable platform for high-gradient THz-driven acceleration, combining simple fabrication and design, strong field enhancement, and compatibility with existing electron sources. It opens new pathways toward practical, tabletop accelerators for scientific and industrial applications.

\end{abstract}

\keywords{Acceleration \and Terahertz \and Simulations \and Waveguide \and DLA \and Photonic}

\section{Introduction}

Particle accelerators have become indispensable tools in various fields, from fundamental research in particle physics to applied sciences and medical therapies \cite{barbalat1990applications,kutsaev2021advanced}. However, the widespread adoption and further development of particle accelerators face significant challenges. Traditional accelerators, with kilometer-scale installations, are restricted not only by their immense size but also by the finite amount of power that can be injected into them to avoid arcing, which can damage the accelerator structure and impair its functionality \cite{hassanein2006effects}.

The scientific community has turned its attention to more compact, high-gradient alternatives in response to these limitations. Dielectric laser-driven accelerators (DLAs) promise to revolutionize the field by dramatically reducing the size of accelerators with acceleration gradients up to two orders of magnitude higher than those of conventional radio-frequency (RF) accelerators \cite{england2014dielectric,peralta2013demonstration}. Despite these promising advances, several unresolved issues persist, notably the difficulties associated with the scaling of energy gain and the challenge of maintaining beam quality over extended acceleration distances.

\begin{figure}[!htb]
    \centering
    \includegraphics[width= 0.9\linewidth]{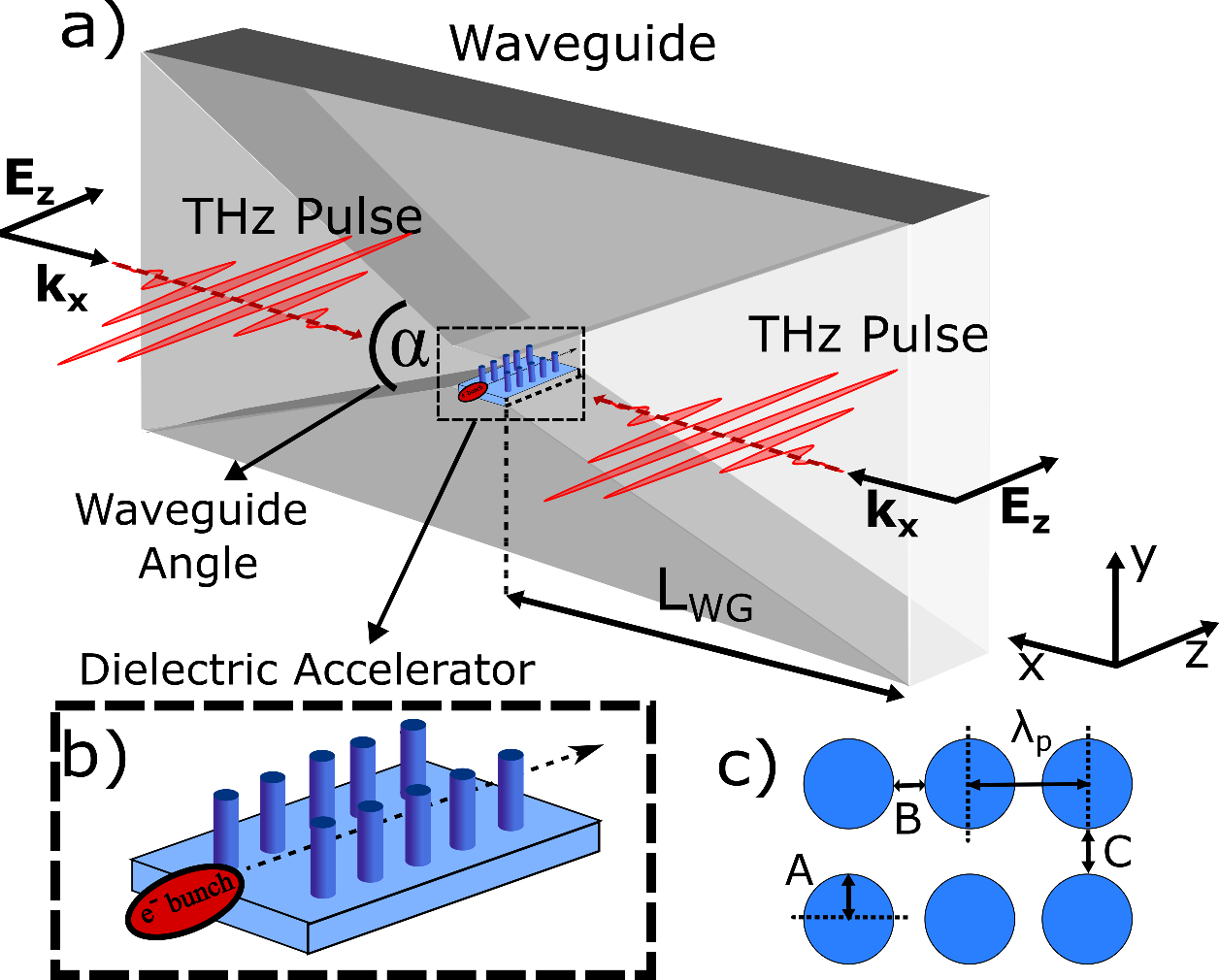}
    \caption{\textbf{a)} Schematic view of the accelerating structure. It operates at a central frequency of $f_{\text{THz}} = \SI{0.65}{\THz} \hspace{0.1cm} (\lambda_{\text{THz}} \approx \SI{461}{\um})$. The proposed setup consists of a symmetric tapered parallel-plate waveguide (TPPWG) of length $L_{\text{WG}} = \qty{27}{\mm}$ and taper angle $\alpha = \ang{14}$ with an integrated \textbf{b)} dielectric THz-driven accelerator (DTA) for electron acceleration. The dual-pillar grating structure dimensions are \textbf{c)} 
    pillar radius $A = \num{0.35} \hspace{0.05cm} \lambda_{\text{THz}}$, pillar trench (inter-pillar distance) $B = \num{0.3} \hspace{0.05cm} \lambda_{\text{THz}}$, pillar height $\qty{400}{\um}$, vacuum channel gap $C = \num{0.4} \hspace{0.05cm} \lambda_{\text{THz}}$, and grating period $\lambda_{\text{p}} = \lambda_{\text{THz}}$. The image is for illustrative purposes only and not to scale.} 
    \label{fig:dta_wg}
\end{figure}

A novel approach to circumvent these drawbacks involves the use of lasers operating at longer wavelengths, such as terahertz (THz) pulses, instead of optical lasers. THz-driven acceleration offers a solution to the limitations faced by optical laser-driven systems \cite{nanni2015terahertz}. The potential of THz-driven acceleration and manipulation is remarkable, offering a host of compelling advantages. Firstly, it allows for a more compact and economically feasible setup that is pivotal for developing high-energy, bright electron beams in a more accessible manner. Secondly, it supports whole-bunch acceleration of relativistic beams with bunch lengths on the picosecond scale and charges reaching tens to hundreds of picocoulombs. Furthermore, metallic waveguides intended to operate in the THz range can already be fabricated with current machining techniques \cite{nanni2015terahertz}. Dielectric microstructures operated in the THz range also offer higher charges per bunch than optical DLAs, and overcome the synchronization issues present in the optical range requiring sub-femtosecond timing jitter. Finally, the scalability and the potential for preserving beam qualities through THz-driven acceleration and manipulation represent a critical advancement towards compact, high-energy THz-driven accelerators. Although single-cycle THz pulses can also be effectively used for particle acceleration \cite{fallahi2016short}, intense, narrowband multi-cycle terahertz pulses are preferred for particle acceleration \cite{fakhari2017thz}. Dielectric terahertz-driven accelerators (DTAs) have the potential to address the critical challenges faced by current accelerator technologies, paving the way for a new generation of particle accelerators that are more compact, scalable, and capable of delivering high-quality beams for a wide range of applications.

The present article proposes a symmetric tapered parallel-plate waveguide (TPPWG) with an integrated dual-pillar grating dielectric structure for particle acceleration operated in the THz range (Figure \ref{fig:dta_wg}). Two multi-cycle linearly polarized terahertz pulses are focused laterally onto the waveguides. The pulses propagate through the waveguides until they reach the DTA pillars, which provide the means for net acceleration by near-fields. Electrons are injected perpendicular to the waveguide along the direction of the dielectric accelerating structure. Particle acceleration can be attained by synchronizing the coupling of THz pulses and the electron injection. THz waveguide studies date back to 2000 \cite{gallot2000terahertz} to analyze the propagation of THz pulses in metallic waveguides. In the following decades, several THz waveguides, their characteristics and properties were studied to enable different applications \cite{atakaramians2013terahertz,andrews2014microstructured}. Different types of waveguides can be employed for particle acceleration, namely hollow-core waveguides, microstructured waveguides, and flat waveguides. From a particle acceleration point of view, waveguides can be divided into co-propagating structure, when the THz pulse propagates in the same direction as the accelerated particle, and perpendicular, when the THz pulse is coupled perpendicular to the acceleration path. Co-propagating waveguides (hollow-core and some microstructured waveguides) have the advantage of longer interaction lengths, and are easier to scale and stage acceleration. On the other hand, very narrow-band pulses are needed, the aperture is small for high-gradient structures, and the walk-off length between the bunch and the wave must be addressed. Perpendicular setups are easier to couple, allow single- to few-cycle pulses to be used without problems, and generally offer larger apertures, but staging is cumbersome and the interaction length is limited. A review of the different types of THz waveguides can be found in \cite{atakaramians2013terahertz,andrews2014microstructured,wu2025review}. Because acceleration is achieved using a dual-pillar grating structure operated in the THz range, a TPPWG integrates quite well with the dielectric accelerator (see section \ref{sec:dielectric_optimization}).
To the extent of the authors' knowledge, no setup combining a waveguide with a DLA has been proposed in the literature. Multiple studies on the use of waveguides and cavities \cite{fallahi2016short,fakhari2017thz,zhang2020cascaded,walsh2017demonstration,zhao2020femtosecond,curry2018meter,wong2013compact,turnar2022waveguide,zhang2018segmented,apsimon2021six,huang2016terahertz,yu2023megaelectronvolt,ying2024high,nix2024terahertz} and dielectric terahertz-driven accelerators \cite{wei2018investigations,xiriai2019numerical,liu2021thz,xiriai2024numerical} for particle acceleration and manipulation have been extensively analyzed with promising results, but without merging both elements into a single accelerating structure. In this work, a multi-cycle THz pulse was generated by a periodically-poled lithium niobate (PPLN) wafer stack. The propagation and enhancement of the THz pulse inside the waveguide are predicted by simulations and investigated experimentally. The pulse is focused onto a homemade antisymmetric waveguide. The outgoing THz pulse (i.e., after traversing the waveguide) temporal profile was measured by electro-optic sampling (EOS). The simulated and measured pulses are compared. After this stage, all subsequent analyses were carried out numerically. The optimizations of the waveguide and dielectric pillars are performed with time-domain simulations. Lastly, the accelerating structure is tested using particle-in-cell simulations, and a few words are given on the THz-induced damage threshold, a key figure of merit for these kinds of accelerators.

\section{Experimental results and simulation validation}

\begin{figure}[!htb]
    \centering
    \includegraphics[width= 0.9\linewidth]{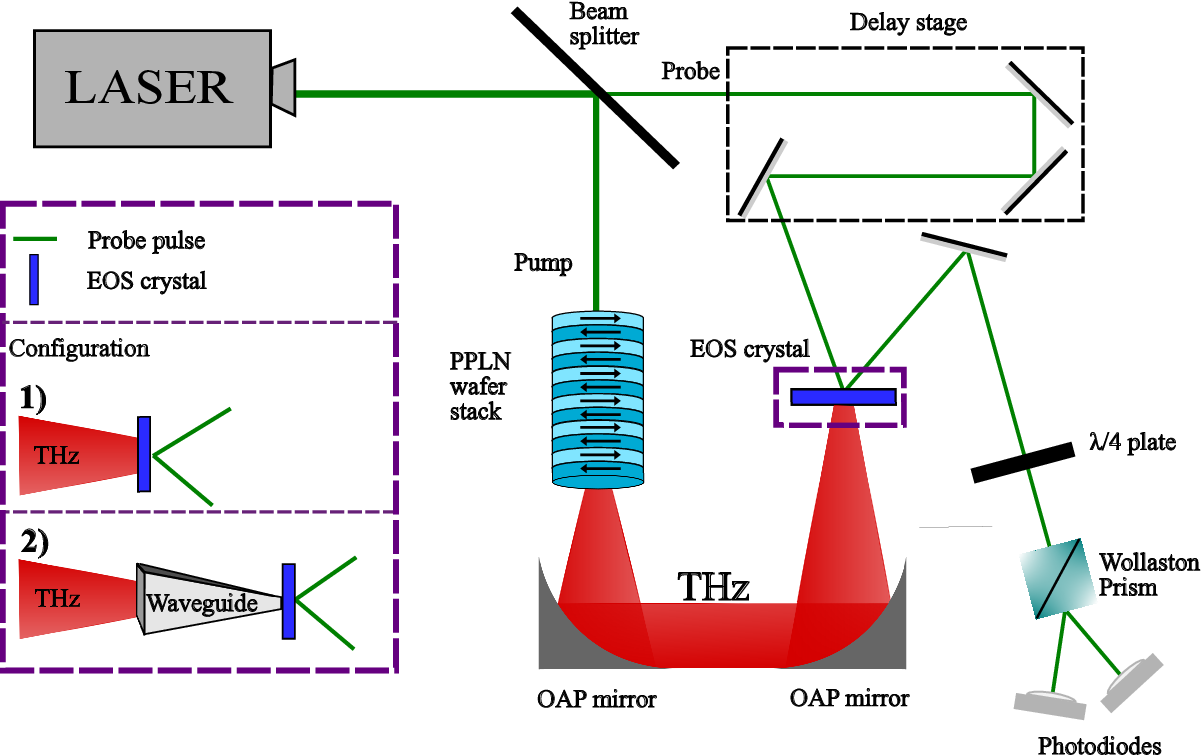}
    \caption{Schematic view of the experimental setup.}
    \label{fig:exp_setup_w_conf}
\end{figure}

Figure \ref{fig:exp_setup_w_conf} depicts the experimental setup used to measure the temporal profile of the electric field and study the effects of the TPPWG on the terahertz pulse. Since it is not possible to measure the field inside the waveguide, an asymmetric version serves for experimental measurement of the THz field to validate the simulation results. All the measurements were performed with no dielectric structure integrated to the waveguide. The multi-cycle THz pulse is generated by optical rectification of an ultra-short, high-intensity optical pulse interacting with a quasi-phase-matched (QPM) nonlinear crystal. The experimental setup consists of a PPLN wafer stack to generate a multi-cycle terahertz pulse with central frequency around \SI{0.65}{\tera \hertz}, followed by a reflective-EOS setup to acquire the THz pulse temporal shape. A \SI{1}{\milli \joule}, \SI{1}{\kilo \hertz} repetition rate, \SI{1030}{\nano \metre} laser is split into a pump and probe. The pump is directed onto the wafer stack to produce the terahertz radiation, while the probe passes through a delay stage and samples the electric field at different time instants. Measurements were performed in two different configurations, shown inside the dashed purple inset. The configurations are: 1) without waveguide, where the terahertz field is focused directly at the EOS crystal (typical EOS setups), and 2) with the waveguide, where the THz pulse is focused onto the waveguide entrance plane. For the second configuration, due to technical constraints, the EO crystal is located \SI{3}{\milli \meter} after the waveguide exit.

\begin{figure}[!htbp]
    \centering
    \includegraphics[width=1\linewidth]{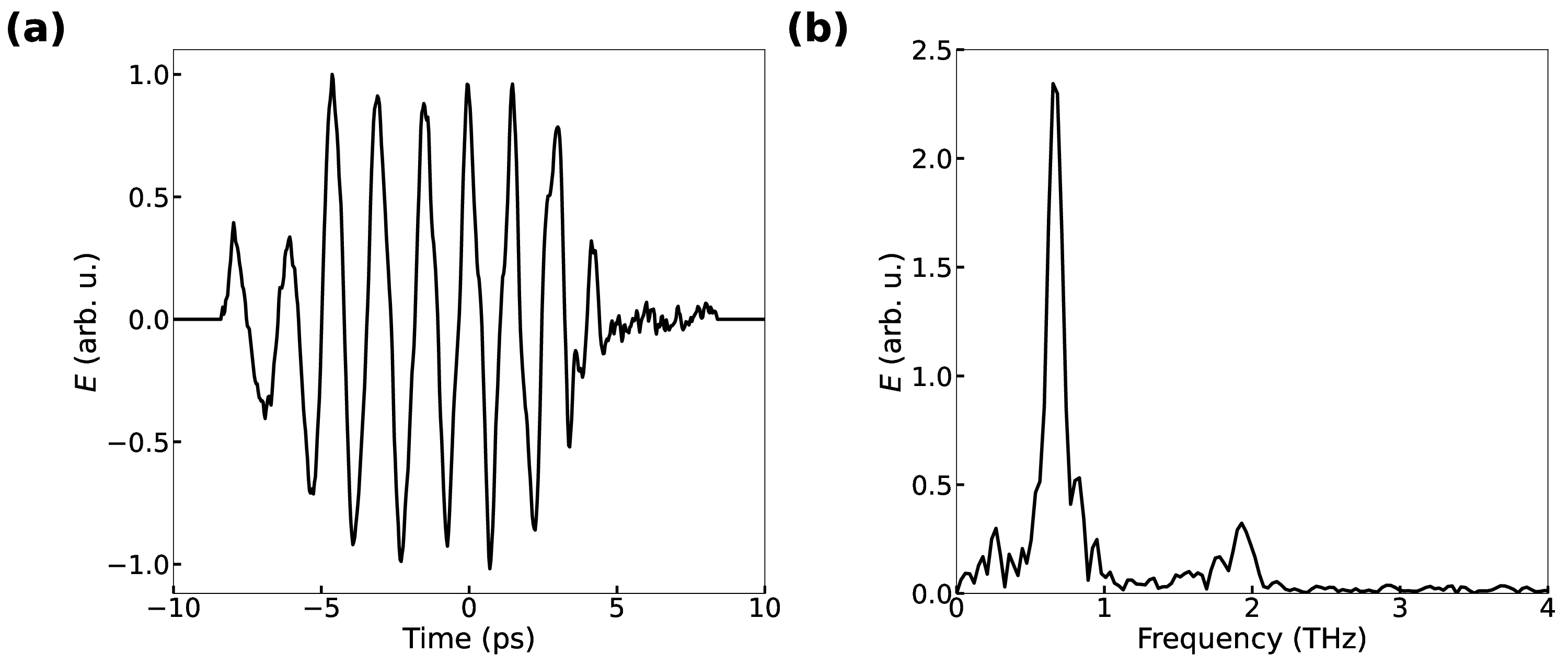}
    \caption{THz pulse \textbf{a)} temporal profile and \textbf{b)} spectrum generated by the PPLN wafer stack. The central frequency is estimated to be around 0.65 THz.}
    \label{fig:thz_pulse}
\end{figure}

The measured THz pulse in the time and frequency domains for the first configuration is shown in Figure \ref{fig:thz_pulse}. The generated THz pulse has 6 main cycles, and a central frequency around \SI{0.65}{\THz}. The number of cycles and the measured frequency are in agreement with the six pairs of wafers used in the PPLN wafer stack with thickness of \SI{80}{\um}. The measured THz temporal profile is then used as the input signal at the waveguide's entrance for all the simulations in this article.

\begin{figure}[!htb]
    \centering
    \includegraphics[scale=0.35]{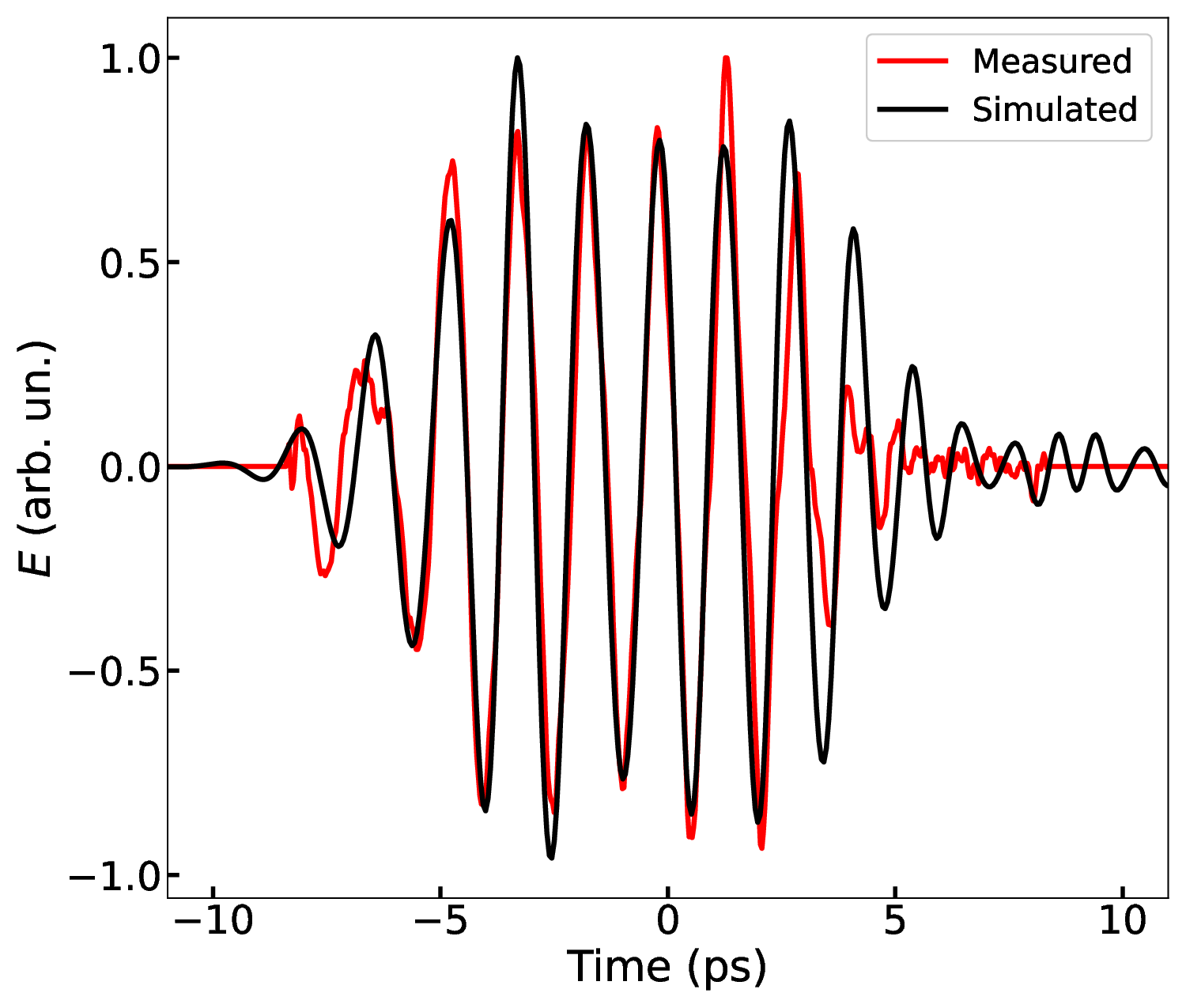}
    \caption{Measured and simulated THz field reaching the EOS crystal for configuration 2). The resemblance between the pulses confirms that simulations can be used for field-waveguide interaction optimization.}
    \label{fig:simu_exp}
\end{figure}

A finite integration technique (FIT) simulation emulating the second configuration was run on CST Studio Suite \cite{studio2010cst} to compare with the experimental results. The measured signal for configuration 1) was used as the excitation signal in the simulation. The measured field in configuration 2) and its simulated counterpart (the computational field calculated at the position where the detector is placed) are compared in Figure \ref{fig:simu_exp}. The similarity of the experimental and simulation pulse shapes validates the simulation accuracy of the THz pulse temporal evolution inside the waveguide, replicating the experimental results to a reasonable extent. Therefore, the excitation pulse propagation and waveguide optimization can be analyzed through simulations. 

\section{Waveguide}

THz waveguides and fibers are paramount for a handful of applications. Naturally, their development has been a cornerstone as THz sources have become available over the past decades. They can be tailored and engineered for specific applications. A review of the different types of THz waveguides and fibers and their corresponding characteristics and applications can be found in \cite{atakaramians2013terahertz,andrews2014microstructured,wu2025review}. In the current study, a waveguide is employed to enhance particle acceleration, i.e. to maximize the electron energy gain. To begin with, a TPPWG helps with the free-space-waveguide impedance matching. Both coupling efficiencies and reflections in the THz range were quantified in \cite{kim2010improvement, mbonye2012study}. For dominant propagating modes, parallel-plate waveguides enable broadband, low-dispersion propagation, as shown in representative works 
\cite{gallot2000terahertz}. Higher-order mode excitation and radiation leakage may occur depending on the waveguide geometry and excitation conditions, as reported in previous research \cite{mueckstein2013mode}. Therefore, a TPPWG aligns accordingly with the purposes of the present work. Introducing a TPPWG has two primary purposes: 1) to couple the THz pulses into the structure \cite{gallot2000terahertz,kim2010improvement,mbonye2012study,othman2019parallel} and deliver them to the dielectric terahertz-driven accelerator, and 2) to amplify the THz electric field \cite{turnar2022waveguide,huang2016terahertz,iwaszczuk2012terahertz}. For a given beam spot size, the TPPWG is capable of coupling the THz pulses with minimal reflection. The terahertz pulse is incident on the waveguide entrance. A guided fundamental excitation at the waveguide entrance reference plane is assumed. It propagates along the tapered horn, increasing the electric field by confining the electromagnetic energy in smaller transversal sections. Therefore, the THz field amplification provided by the TPPWG boosts particle acceleration by increasing the electric field in the dielectric accelerator.

\begin{figure}[!htb]
    \centering
    \subfloat{\includegraphics[width=0.5\linewidth]{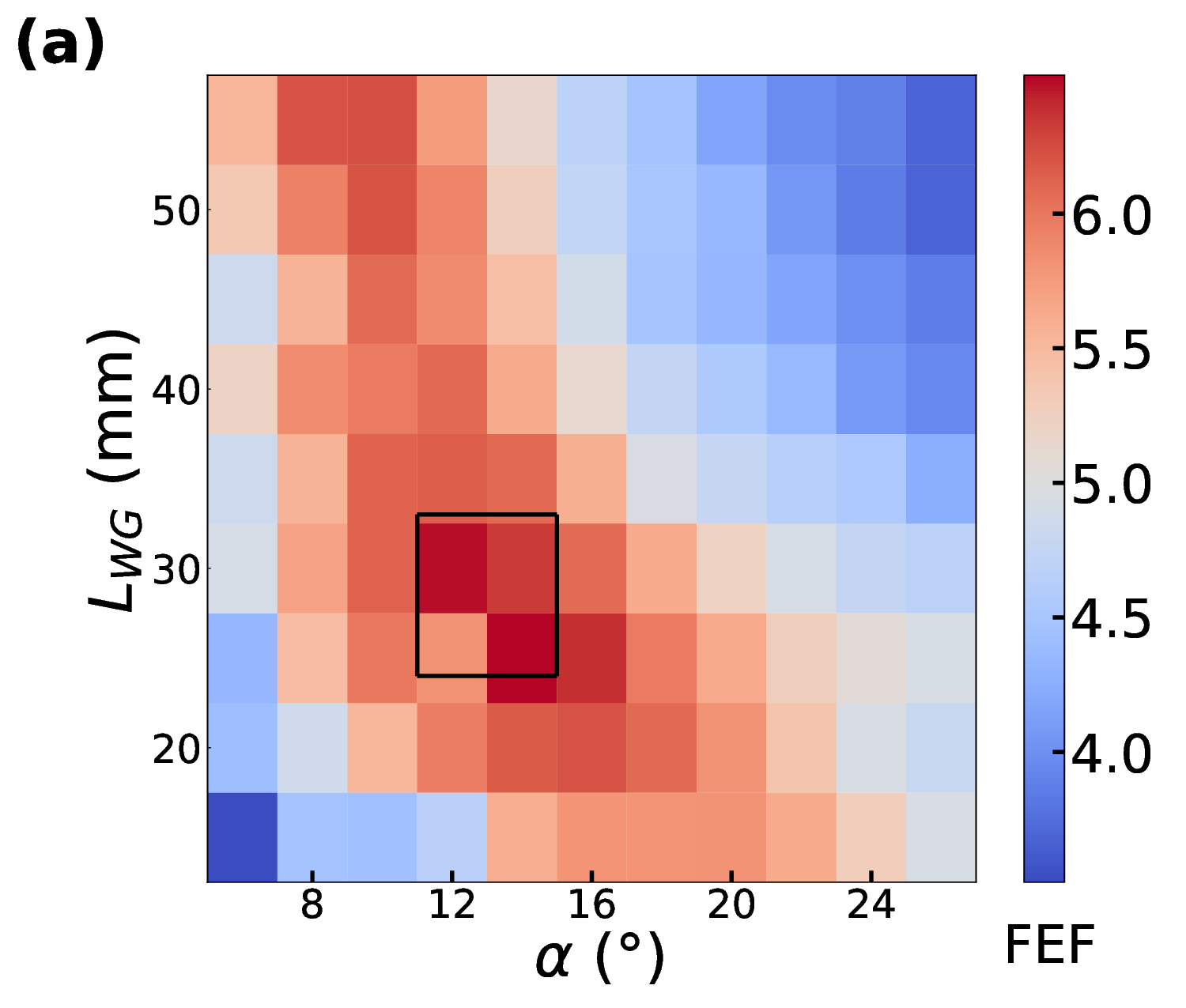}}\hfill
    \subfloat{\includegraphics[width=0.5\linewidth]{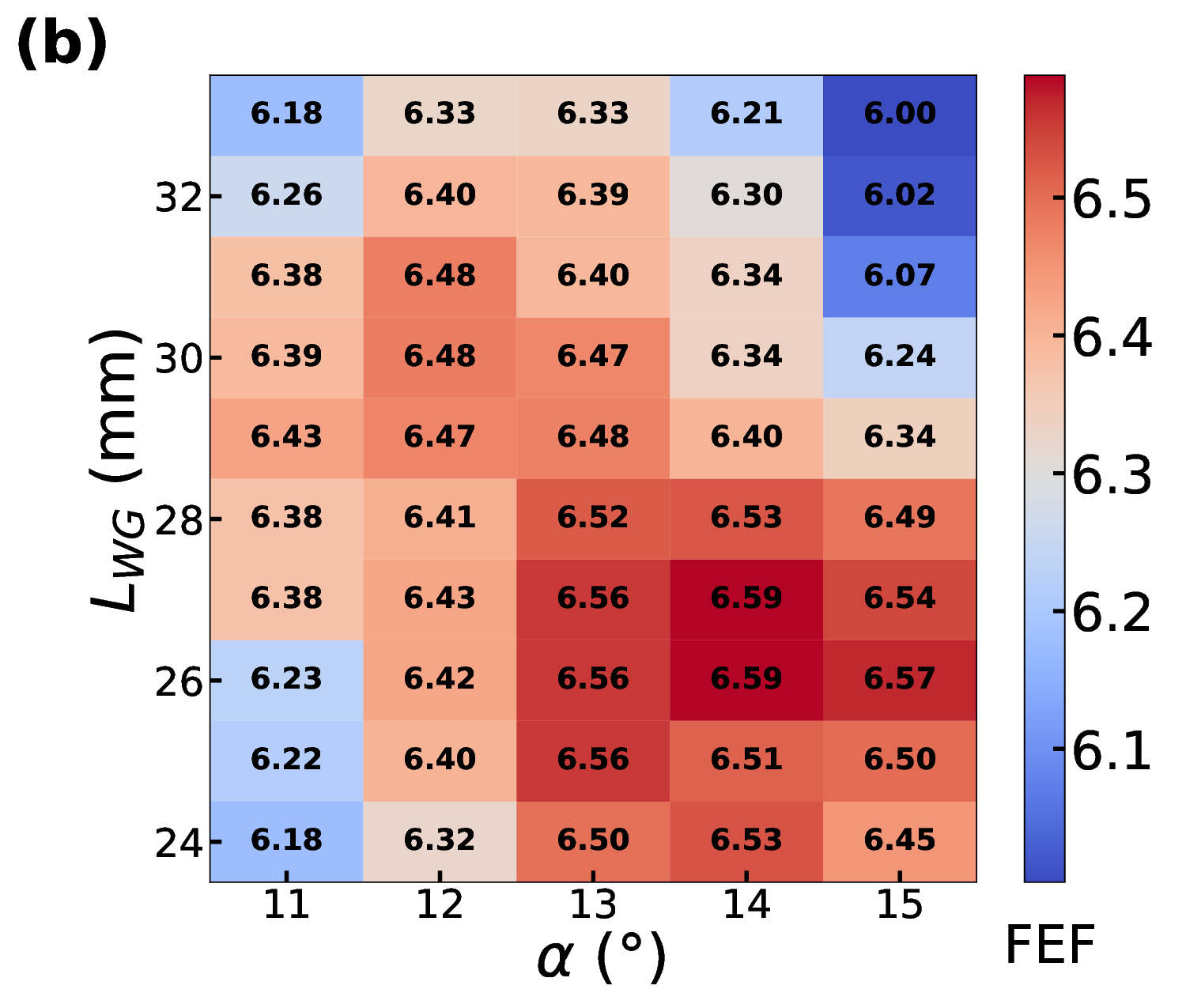}}
    \caption{\textbf{a)} represents the waveguide field enhancement factor (FEF) and \textbf{b)} it is a zoom-in version according to the structural parameters length $L_{\text{WG}}$ and angle $\alpha$. A higher-resolution parametric sweep was performed inside the black inset in Figure (a) to find the optimal length and angle values that maximize the electric field.}    
    \label{fig:AmpFac}
\end{figure}

Indeed, the waveguide structural parameters directly affect the THz pulse intensity and, by extension, the electron energy gain. The waveguide exit gap (i.e. the distance between the plates at the waveguide output/end facet) is \qty{400}{\um} to match the pillars height(refer to section \ref{sec:dielectric_optimization}). Waveguide optimization (i.e., maximizing the electric field reaching the dielectric structure) is based on the waveguide length $L_{\text{WG}}$ and angle $\alpha$. A parametric sweep varying the waveguide length (from 10 to 60 mm, in steps of 5 mm) and angle (from 6 to 26 degrees, in steps of 2 degrees) was performed in CST MWS (FIT) to calculate the electric field inside the structure. Figure \ref{fig:AmpFac}a) color map shows the electric field enhancement factor (FEF) as a function of the waveguide length and angle. The FEF is calculated as the ratio between the electric field peak value at the pillars' position and the entrance of the waveguide ($f_{\text{E}} = E_{\text{max}}^{\text{DLA}} / E_{\text{max}}^{\text{Entrance}}$). The FEF maximum value is found for a waveguide length of 25 mm and a flare angle of 14 degrees. A second parametric sweep was performed around these values for fine-tuning. The simulated values for the length and angle were 24-33 mm and 11-15 degrees, respectively. The results can be seen in Figure \ref{fig:AmpFac}b). The maximum value is achieved for 14 degrees and 27 mm for the angle and length, respectively. The field enhancement $f_{\text{E}}$ does not change significantly around these values, which is advantageous from a practical standpoint. The tolerances introduced by the manufacturing process will not substantially affect the FEF of the prospective accelerator.



\begin{figure}[!htb]
    \centering
    \includegraphics[width=0.90\linewidth]{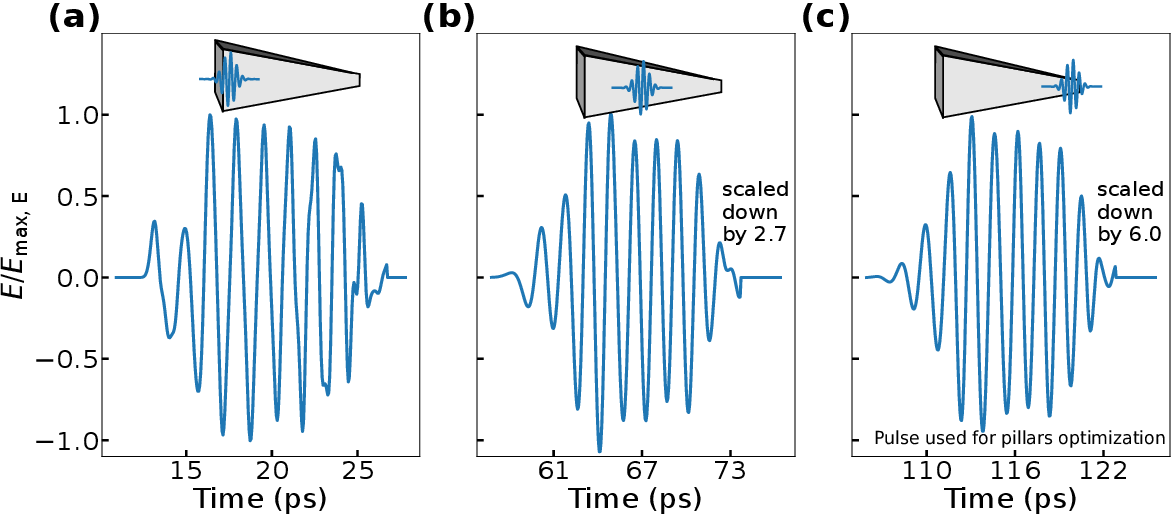}
    \caption{The THz pulse normalized to the entrance peak field ($E_{\text{max, E}}$) at \textbf{a)} the waveguide entrance plane, \textbf{b)} the waveguide half-length, scaled down by a factor of 2.7, and \textbf{c)} before the waveguide exit plane, where the micro-accelerator is meant to be placed, scaled down by a factor of 6.0. This pulse shape \textbf{c)} was used for the pillar optimization (see section \ref{sec:dielectric_optimization}).}
    \label{fig:pulse_propagation_w_wg}
\end{figure}

Figure \ref{fig:pulse_propagation_w_wg} depicts the THz pulse propagation and the temporal evolution along the TPPWG for $L_{\text{WG}} = \SI{27}{\mm}$ and $\alpha = \SI{14}{\degree}$. The electric field normalized to the input peak signal is shown at different positions inside the waveguide (entrance, middle, and exit). The pulse shape at the waveguide entrance is depicted in Figure \ref{fig:pulse_propagation_w_wg}a). As explained before, the electric field is increased along the tapered section, as portrayed in Figure \ref{fig:pulse_propagation_w_wg}b) and \ref{fig:pulse_propagation_w_wg}c), where the fields were amplified at least by a factor of 2.7 and 6.0, respectively. The time-domain field at the accelerator/pillars plane (Figure \ref{fig:pulse_propagation_w_wg}c)) is the THz waveform input used for the dielectric optimization analysis (see section \ref{sec:dielectric_optimization}).


\section{Dielectric structure optimization}
\label{sec:dielectric_optimization}

Several geometries and elements have been proposed for DLAs, namely photonic crystals \cite{lin2001photonic,smirnova2004photonic,locatelli2017photonic}, distributed Bragg reflectors (DBR) \cite{mizhari2004optical}, grating structures \cite{peralta2013demonstration, plettner2006proposed}, and inverse Cherenkov radiation (ICR) \cite{liu2020microscale}, each one of them with advantages and disadvantages. All of these types of dielectric microstructures can be operated in the THz range, with minimal modifications \cite{wei2018investigations,xiriai2019numerical,liu2021thz,palfalvi2014evanescent}. The DTA selected for this work is a cylindrically shaped dual-pillar grating structure \cite{leedle2015dielectric}. Grating structures provide high-gradient acceleration with minimal to no deflection \cite{breuer2014dielectric}, transverse focusing can be achieved with alternating-phase focusing (APF) \cite{niedermayer2018alternating,chlouba2023coherent}, and they can be manufactured with relative ease with current techniques in material science and can be tailored for different aims (acceleration, focusing, deflection, etc) \cite{yousefi2018silicon}. Also, dual-pillar grating structure performance improves when supplied symmetrically with two lasers from opposite sides \cite{leedle2015dielectric}, which synergizes smoothly with the TPPWG. In addition to this, the dual-pillar gratings are less prone to fabrication tolerances \cite{leedle2015dielectric,yousefi2018silicon} and can be integrated and scaled adequately with waveguides and fiber-optic systems. Other configurations can also be implemented for particle acceleration. The proposed setup could be operated by either illuminating a single side (asymmetric TPPWG) or both sides of the waveguide (symmetric TPPWG). For instance, a possible single-side laser operation would consist of a grating structure with a DBR \cite{xiriai2019numerical,yousefi2018silicon} in cases where laser phase control is difficult to implement in practice. Other DTA accelerators could be improved by including the TPPWG in their design. For example, the performance of an ICR-DTA consisting of a single or double prism could be improved because the field enhancement supplied by the TPPWG increases the THz pulse impinging on the prisms, allowing the use of multi-cycle terahertz pulses to decrease the dispersion effects and facilitate its implementation in practice. 


Once the waveguide parameter values are defined, appropriately shaping the dielectric microstructure will also increase the energy delivered to the electron in the acceleration process. Silicon was selected as the pillar material. Its refractive index in the THz range is 3.42, according to \cite{dai2004terahertz}. For relativistic electron acceleration, the grating period must match the laser wavelength $\lambda_p = \lambda_{\text{THz}}$ \cite{breuer2014dielectric}. Optimization of the dielectric grating parameters, such as channel gap $g$ and pillar radius $r_{\text{p}}$, was carried out taking into account the acceleration gradient $G_0$ and the acceleration factor AF, following a similar procedure in \cite{wei2018investigations,xiriai2019numerical,wei2017beam}. The formulas for these DLA figures of merit are \cite{peralta2015accelerator}:

{\centering
  $ \displaystyle
    \begin{aligned} 
    & G_0 = \frac{q}{\lambda_p} \int_{0}^{\lambda_p} E_z(z(t), \phi) dz, \hspace{0.1cm} \text{and} \\
    & \text{AF} = \frac{G_0}{E_{\text{max}}},
    \end{aligned}
  $ 
\par}

respectively.

\begin{figure}[!htb]
    \centering
    \includegraphics[width=0.85\linewidth]{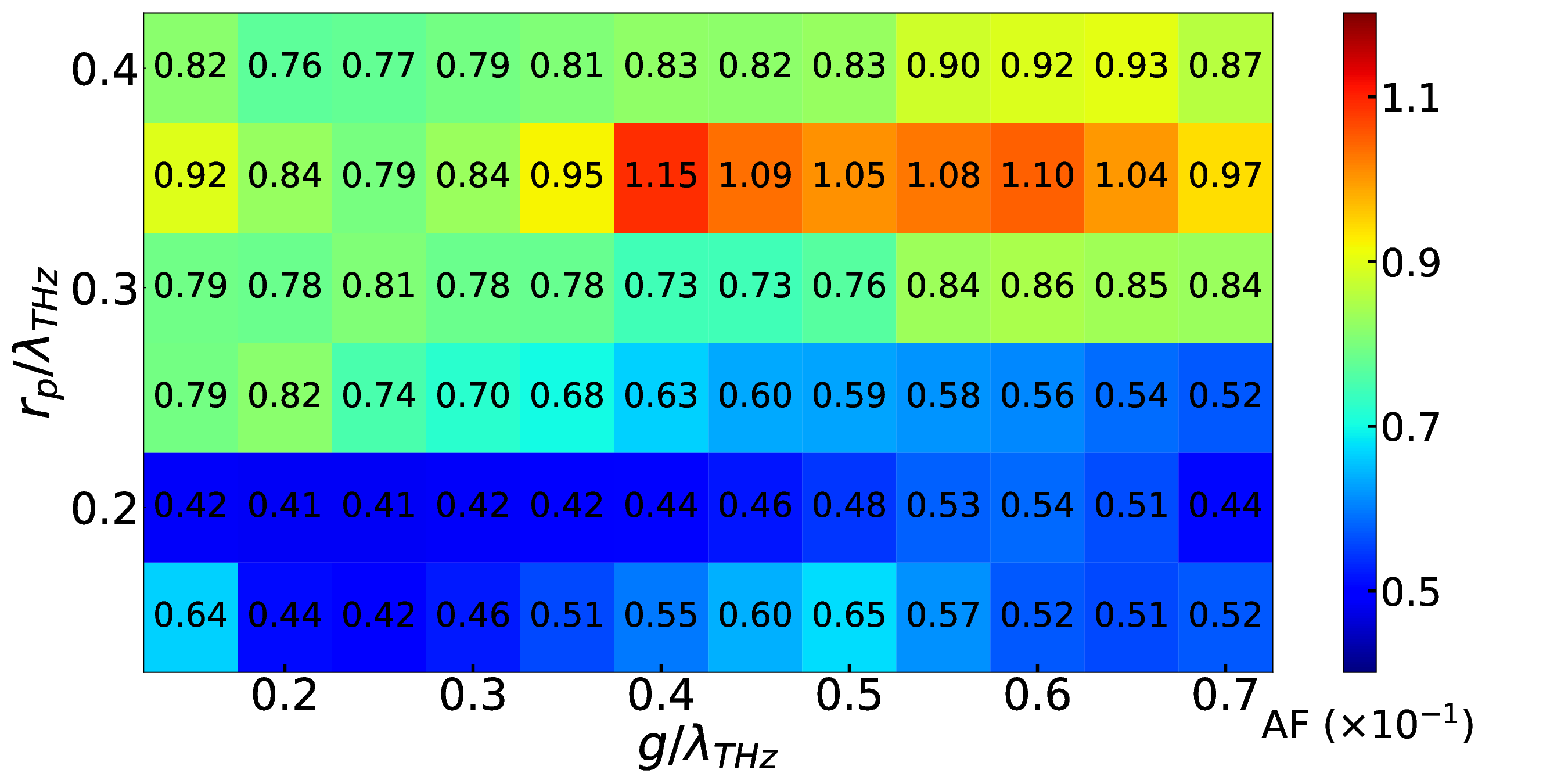}
    \caption{Color mapped data of the acceleration factor for cylindrical pillars in terms of the pillar radii and vacuum channel gap.}
    \label{fig:AF}
\end{figure}

The acceleration gradient is the average longitudinal electric field experienced by an electron injected at the optimal phase $\phi$. The acceleration gradient is calculated in one grating period ($\lambda_p$) for the electron path $z(t)$. Because the pulse shape is somewhat altered as it propagates along the waveguide (refer to Figure \ref{fig:pulse_propagation_w_wg}), this must be considered for the pillar optimization. The pulse temporal shape impinging the pillars was taken from the waveguide simulation evaluated at the DTA pillar position (see Figure \ref{fig:pulse_propagation_w_wg}c)). The accelerating factor was calculated for values of the pillar radius and channel gap ranging between the values $(0.15-0.40) \times \lambda_{\text{THz}}$ and $(0.15-0.70) \times \lambda_{\text{THz}}$, respectively. The results are shown in Figure \ref{fig:AF} as a color map. The energy gain of the electrons depends on the phase mask provided by the dual-pillar grating structure on its pair of values (radius, gap). The AF maximum is found for the pair value $(g, r_{\text{p}}) = (0.40, 0.35) \lambda_{\text{THz}}$. There are several advantages offered by this pair value. First, the gap channel is wide, which softens the bunch transverse size restrictions, may allow higher charge per bunch, and facilitates the transport of the beam, thanks to the extra transversal space. Second, the fabrication errors are lower in longer-radii pillars, increasing the relative fabrication tolerances. According to these results, a DTA-integrated THz waveguide with a 27 mm length and an angle of 14 degrees with a dual-pillar grating structure of silicon with a radius of 0.35$\times \lambda_{\text{THz}}$, pillar height of \SI{400}{\um}, and a vacuum channel gap of 0.40$\times \lambda_{\text{THz}}$ provides one of the highest attainable energy gains for the given configuration.

\pagebreak

\section{Accelerator performance}

The accelerator performance was evaluated by injecting a single pre-bunched electron beam with several different bunch charges. Particle-in-cell (PIC) simulations using CST Studio Suite \cite{studio2010cst} were carried out for the proposed accelerator to study the electron bunch energy gain, energy spread, acceleration gradient, and other important quantities characterizing accelerators. The accelerator performance was calculated by injecting a single, pre-bunched electron beam for several values of the bunch charge. Table \ref{tab:bunch} shows the bunch parameter values for the simulations. These bunch parameters are already available with current accelerator technology \cite{PhysRevAccelBeams.23.044801,snedden2023specificationdesignenergybeam}. Moreover, the achievable values for bunch lengths can be shorter than that used in this work, diminishing the energy spread caused by the acceleration process.

\begin{table}[!htb]
    \centering
    \begin{tabular}{c|c}
        Bunch Parameter & Value \\
        \hline  
        Transverse size  & \SI{50}{\micro \meter}\\
        Bunch length (FWHM) & \SI{250}{\femto \second}\\
        Energy & \SI{6}{\mega \eV} \\
        Energy spread & 1 \% \\
        Angular spread & 1 \% \\
        Emittance & \SI{0.4}{\um} rad \\
        Bunch charge & from \SIrange[]{0.01}{100}{\pico \coulomb} \\
    \end{tabular}
    \caption{Bunch parameters used in the PIC simulations.}
    \label{tab:bunch}
\end{table}

The electron bunch is injected so that the electron energy gain is maximized. The highest peak electric field of the THz pulse used was around \SI{250}{\kV \per \cm}, with a central frequency of \SI{0.65}{\THz} at the waveguide entrance plane. This peak field value can be accomplished with a relatively compact THz source based on a cryogenically cooled-down, large-area PPLN wafer stack pumped with an IR laser with short duration and pulse energy of the order of $\SIrange[]{0.1}{1}{\joule}$. Several techniques, schemes, and novelties on the generation of mJ energy, narrowband, multi-cycle THz pulses with the mentioned setup have seen improvements in recent years \cite{ahr2017narrowband,lemery2020highly,mosley2023large,rentschler2024parameter,dalton2024cryogenically}, and it is more likely that the THz community will reach the energy level demands for particle acceleration in the near future. The acceleration length in the simulations was limited by the large number of cells needed to run a large structure compared to the accelerating wavelength ($2 \times L_{\text{WG}}\approx \SI{55}{\mm} >> \lambda_{\text{THz}} (\approx \SI{461}{\um})$). Acceleration was simulated for a grating with 5 periods, resulting in a total accelerator length of $5\lambda_{\text{THz}} \approx \SI{2.3}{\mm}$. 

In the case of ultra-relativistic beams, on-crest acceleration can be used to maximize the energy gain of the bunch without a significant amount of phase slippage. The phase slippage can be calculated in the following way: 

\begin{equation}
    \Delta \phi \approx \frac{\omega}{c} \int_{0}^{z} \frac{\Delta \beta}{\beta^2_s(z')} dz' = \frac{2 \pi}{\lambda} \int_0^{z} \frac{\Delta \gamma}{\gamma_s^{3}(z') \beta_s^{3}(z')} dz'
\end{equation}

The phase slippage over the accelerator length is estimated to be around 0.11 \%. More critical for on-crest acceleration is the value of the bunch length. Short bunch lengths are desired to minimize the energy spread during acceleration. In this case, the bunch (FWHM) occupies around $0.163 \times T_{\text{THz}}$, with $T_{\text{THz}}$ being the period of the accelerating wave. The energy gain between the center and the head/tail of the bunch is different. Given the relative high percentage of the bunch length compared with the wave period, the energy spread increases during acceleration. The energy spread can be minimized by using shorter bunch lengths or by accelerating the bunch slightly off-crest to imprint a linear energy chirp and use a magnetic chicane for compression.

\begin{figure}[!htb]
    \centering
    \subfloat{\includegraphics[width=0.5\linewidth]{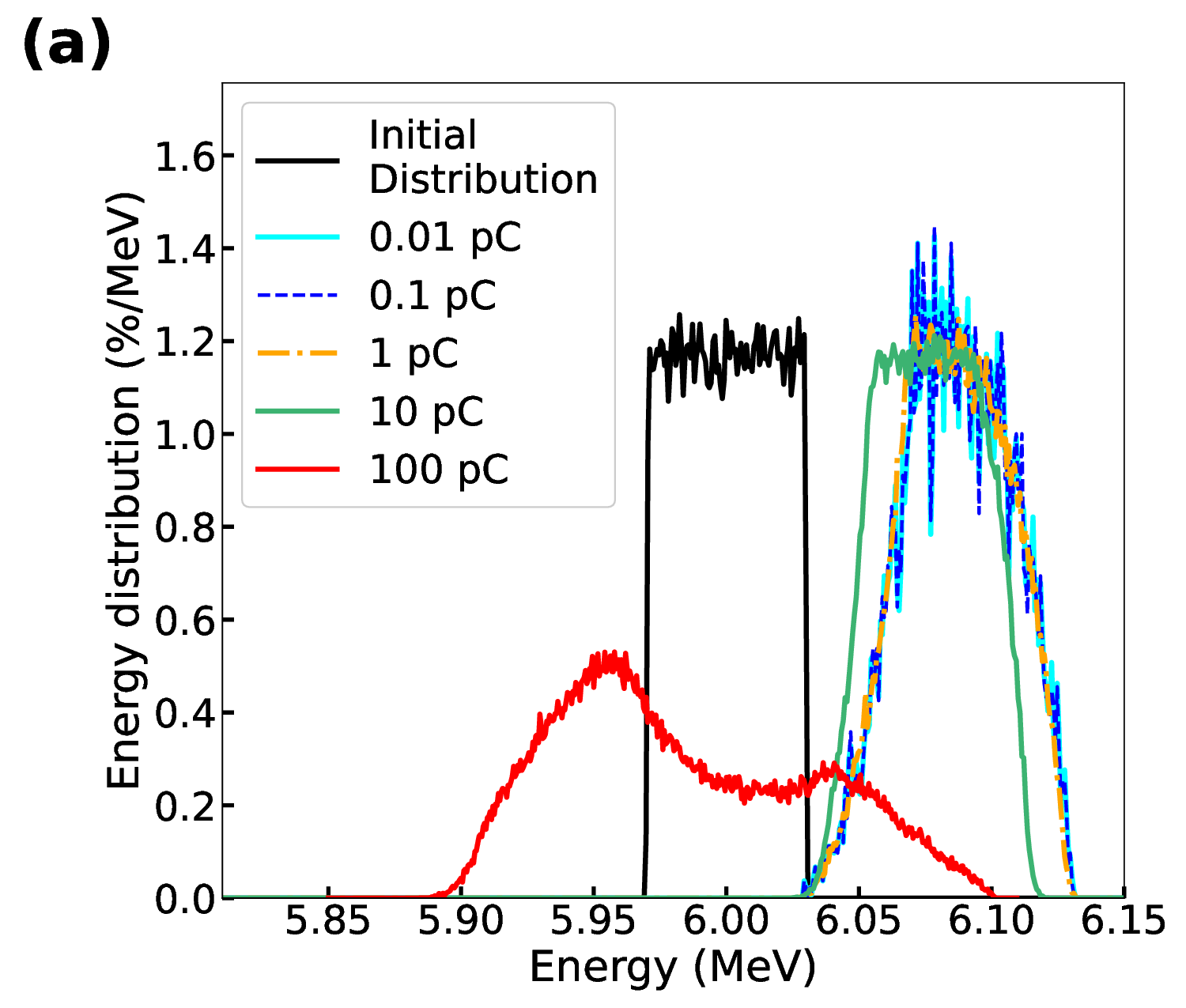}}\hfill
    \subfloat{\includegraphics[width=0.5\linewidth]{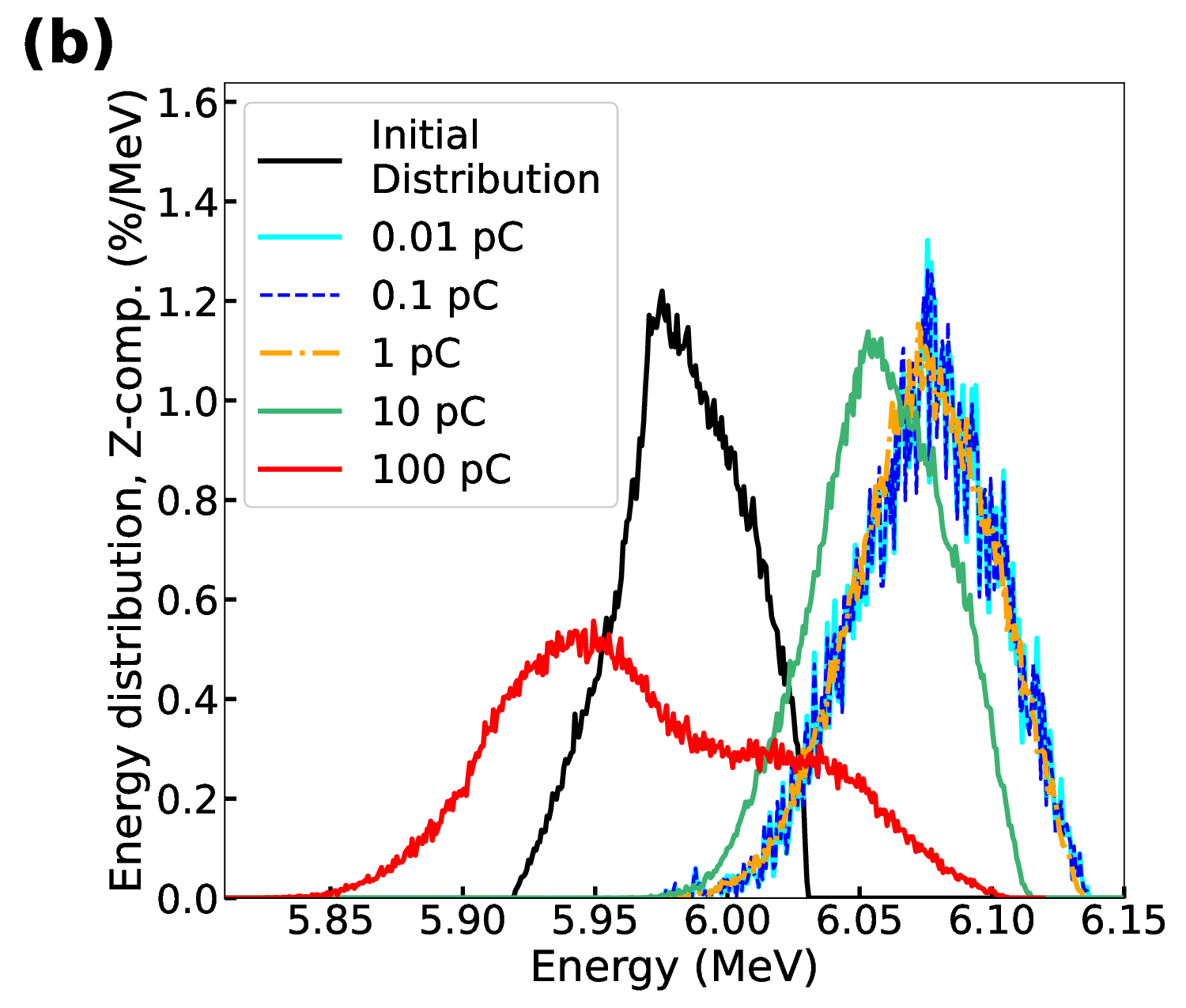}}
    \caption{The energy distribution for the initial bunch and the post-accelerated bunch for different bunch charges. Figure \textbf{(a)} represents the total kinetic energy distribution while \textbf{(b)} considers the kinetic energy projected along the accelerating path (z-direction).}
    \label{fig:energy_distribution}
\end{figure}

The kinetic energy and its projection along the accelerating path (calculated by $E^{\text{kin}}_{\text{z}} = (\gamma_{\text{z}} - 1)mc^{2}$, where $\gamma_{\text{z}} = 1/\sqrt{1 - \beta_{\text{z}}^{2}}$ is the relativistic factor along the accelerating path) are shown for a waveguide with angle $\alpha = \SI{14}{\degree}$ and length $L_{\text{WG}} = \SI{27}{\mm}$, and DTA dual-pillar grating microstructure with pillar radius $r_{p}=0.35\lambda_{\text{THz}}$ and vacuum channel gap $g = 0.4\lambda_{\text{THz}}$ in Figure \ref{fig:energy_distribution} for several values of the bunch charge. Evidently, the energy distribution does not change significantly for bunch charges between \SI{0.01}{\pico \coulomb} and \SI{1}{\pico \coulomb}. In the case of the projected kinetic energy, a tail can be observed in the energy distributions. This is due to the fact that the initial bunch has an angular distribution. Consequently, the kinetic energy of some electrons along the z-axis is lower than the initial energy. For a bunch charge of \SI{10}{\pico \coulomb}, the median of the projected kinetic energy is approximately \SI{6.05}{\MeV}, in which case, the space charge effects begin to disrupt the acceleration process of the bunch as a whole. A different distribution can be observed for a bunch with total charge equal to $\SI{100}{\pico \coulomb}$. For values close to the latter, the space charge effects dominate the interaction in the bunch according to our simulations. The projected kinetic energy shows no trend along the direction of acceleration, and instead, the mean energy of the electron bunch turns out to be lower than the initial energy. The electrons deviate from the main axis considerably, compromising the integrity of the bunch. 


\begin{figure}[!htb]
    \centering
    \includegraphics[width= 0.95\linewidth]{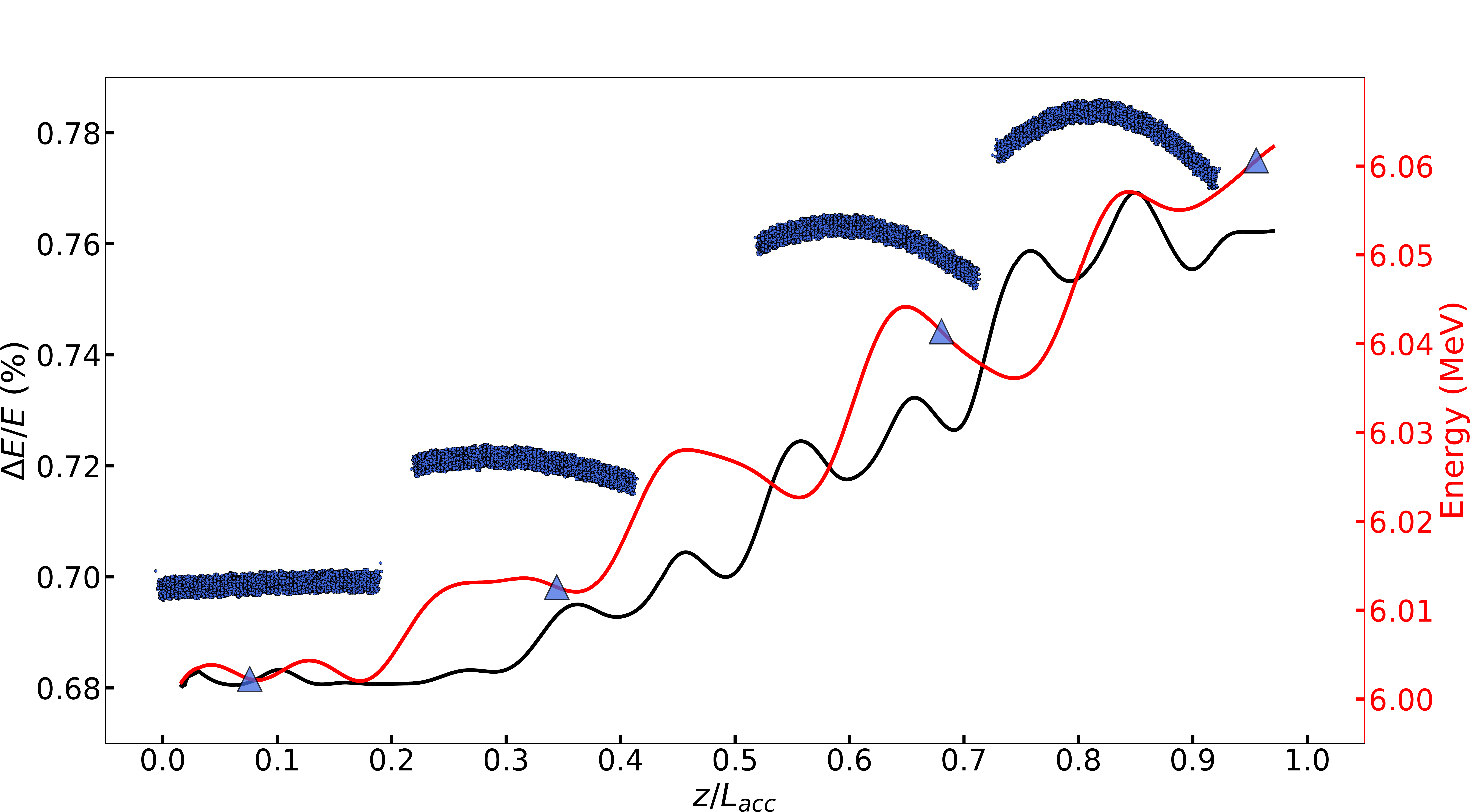}
    \caption{Bunch energy spread (black curve) and energy evolution (red curve) along the accelerator. Also, four selected snapshots of the bunch longitudinal phase space are illustrated at the following positions approximately: $(0; \hspace{0.1cm} 1/3; \hspace{0.1cm} 2/3; \hspace{0.1cm} 1) \times L_{\text{acc}}$, featured by the blue triangles.}
    \label{fig:energyPos}
\end{figure}

Figure \ref{fig:energyPos} portrays the bunch kinetic energy and energy spread along the accelerator for \SI{1}{\pico \coulomb} bunch charge. In addition, four snapshots of the bunch distribution are shown at different positions along $z$. Initially, the bunch energy spread is given only by the initial energy spread and angular spread of the bunch. Right after injection, the energy spread increases because electrons at different positions within the bunch experience different electric field values, causing different energy gains inside the bunch. This is because the bunch length (FWHM) is around one-sixth of the accelerating field's wavelength ( bunch length $\approx \SI{75}{\um}$ and $\lambda_{\text{THz}} \approx \SI{461.215}{\um}$). The energy spread becomes larger as the bunch advances along the accelerator. It can also be noted that the bunch energy is not a monotonically increasing function of the position, but the bunch gains and loses energy along the acceleration process. This is in agreement with the DLA grating acceleration concept, where the electrons are continuously accelerated and decelerated, but because acceleration is stronger than deceleration, a net positive energy gain is observed. On the other hand, the bunch length remains unchanged during the acceleration, as expected for an ultra-relativistic electron bunch. 

Finally, the acceleration gradient was calculated for a higher peak electric field of $\SI{1}{\mega \volt \per \cm}$. These field values are one order of magnitude less than those required for particle accelerators. All the other parameters remained unchanged. An energy gain of $\SI{0.276}{\MeV}$ was obtained in a path length of \SI{0.23}{\cm}, resulting in an acceleration gradient of \SI{1.2}{\MeV \per \cm}. High values of input electric fields ($\sim \SIrange[]{10}{100}{\mega \volt \per \cm}$) induce large electric fields inside both the metallic waveguide and the dielectric microstructure. In this case, the electric field values should be adjusted to the materials' THz-induced damage threshold to prevent breaking the structure. The first study towards THz-induced damage threshold in dielectrics (fused silica) was done by \cite{thompson2008breakdown}, reporting a \SI{138 \pm 7}{\mega \volt \per \cm} damage threshold. In another study, no apparent damage was found on the n-doped silicon wafers when illuminated with few-cycles, sub-picoseconds THz fields up to \SI{3.1}{\mega \volt \per \cm} \cite{chefonov2017giant}. A similar case is found in \cite{tarekegne2017impact}, which employed THz peak electric fields of \SI{3.6}{\mega \volt \per \cm} in silicon without damage of the structure. The conductivity induced by high-field THz waves in dielectric wakefield accelerators (DWA) is mentioned in \cite{o2019conductivity}. In the article, wave attenuation around \SI{8.5}{\mega \volt \per \cm} is reported for silica. This value could be used as a practical damage threshold for high-field terahertz pulses. Regarding metallic structures, the authors in \cite{zhu2016robustness} analyze the effect of high-intensity THz fields (several \unit{\mega \volt \per \cm}) in hundred-nanometers thick samples of copper, gold, aluminum, and platinum, where clear damage was observed. However, sample thickness (hundreds of nanometers) and local field enhancement might play an important role in these studies. Lastly, several studies provide a similar value for the THz-induced, single-pulse damage threshold for nickel and aluminum, with a field intensity of approximately \SI{15}{\mega \volt \per \cm} \cite{chefonov2018interaction,v2018damage,agranat2018damage}. From these values, this work considers that input fields of the order of \SI{1}{\mega \volt \per \cm} are realistic and practical values for DTA-integrated TPPWG accelerators without impairing their structure. However, the values of the THz-induced  damage threshold in materials are still not well known and further analysis must be carried out for the precise and accurate modeling of THz-induced damage mechanisms. It is beyond the scope of this work to engage in such analysis and shall be reported elsewhere. Typical acceleration gradients of RF cavities range from \SIrange[]{0.5}{1.5}{\MeV \per \cm}. Hence, the TPPWG-DTA operated at field values of \SIrange{5}{10}{\mega \volt \per \cm} provides higher acceleration gradients than current accelerating structures, allowing for more compact setups while decreasing the cost of accelerator facilities.

\section{Conclusion}

In this study, a novel, high-gradient, compact accelerating structure based on a DTA-integrated TPPWG is proposed for particle acceleration driven by multi-cycle terahertz pulses. 
The structure was studied and analyzed through simulations, in addition to an experimental result for comparison. The waveguide structural parameters were shown to have a direct effect on the field amplification. Also, the waveguide can be shaped according to the specific situation and goal. A waveguide length of $L_{\text{WG}} = \SI{27}{\mm}$ and a taper angle of $\alpha = \SI{14}{\degree}$ represent the optimal configuration for maximum amplification within these ranges for an six main cycles THz pulse with central frequency around \SI{0.65}{\THz}.  The TPPWG proves to be a fruitful addition to THz-driven accelerator systems as it helps focus the THz pulses and enhances the field values, mitigating the effects of low optical-to-terahertz conversion efficiencies in current multi-cycle THz generation techniques. The TPPWG synergizes with the dual-pillar gratings as it allows for illumination from both sides, enhancing acceleration. The dielectric structure was also optimized for acceleration by maximizing the AF. The proposed accelerator can support bunch charges around \SI{10}{\pico \coulomb} with input peak fields of the order of $\SI{0.25}{\mega \volt \per \cm}$. These bunch charge values are one to two orders of magnitude lower than those provided by RF cavities. Nevertheless, it is expected to handle higher bunch charges for more energetic THz pulses. For bunch parameters attainable with current accelerators, an energy gain of \SI{87}{\keV} over \SI{2.3}{\mm} was reached with the mentioned parameters. The energy spread increased during the acceleration process; however, this can be addressed by compressors or by providing the structure with shorter bunch lengths. The resulting acceleration gradient was \SI{0.377}{\MeV \per \cm}, while its value increased to \SI{1.20}{\MeV \per \cm} for electric fields of the order of \SI{1}{\mega \volt \per \cm}, matching current radiofrequency cavities. Furthermore, peak fields in the range \SIrange[]{5}{10}{\mega \volt \per \cm} can provide greater acceleration gradients. All things considered, the proposed device is an innovative and compact accelerator that combines two simple elements, a TPPWG and a DTA, for efficient particle acceleration and could lead the way to implementing THz-driven accelerators. Further simulations and experiments are still needed to obtain valuable insights into the DTA-integrated TPPWG accelerator.

\section*{CRediT authorship contribution statement}
\textbf{A. Leiva Genre:} Conceptualization, Formal analysis, Investigation, Methodology, Validation, Visualization, Writing - original draft. \textbf{Z. Tibai:} Conceptualization, Validation, Supervision, Writing - review and editing. \textbf{L. Nasi:} Investigation, Methodology, Validation, Writing - review and editing. \textbf{M. Kiss:} Investigation, Methodology. \textbf{G. Almási:} Conceptualization, Funding acquisition, Project administration, Resources, Supervision, Writing - review and editing. \textbf{J. Hebling:} Funding acquisition, Project administration, Resources, Supervision, Writing - review and editing. \textbf{Sz. Turnár:} Conceptualization, Funding acquisition, Project administration, Resources, Supervision, Validation, Visualization, Writing - review and editing.\\

\section*{Declaration of competing interest}

The authors declare that they have no known competing financial interests or personal relationships that could have appeared to influence the work reported in this paper.

\section*{Data availability}
Data underlying the results presented in this paper are not publicly available at this time but may be obtained from the authors upon reasonable request.

\section*{Funding}
This project has received funding from the European Union's Horizon Europe research and innovation programme under grant agreement no. 101073480 and the UKRI guarantee funds. Also, project no. TKP2021-EGA-17 has been implemented with the support provided by the Ministry of Culture and Innovation of Hungary from the National Research, Development and Innovation Fund, financed under the TKP2021 funding scheme. The project 2024-2.1.1-EKÖP funded by the ministry of culture and innovation, national fund for research, development and innovation, under the University research grant programme EKÖP-24-4 (Szabolcs Turn\'ar).

\bibliographystyle{unsrt}  
\bibliography{sample}  

@techreport{barbalat1990applications,
  title={Applications of particle accelerators},
  author={Barbalat, Oscar},
  year={1990},
  institution={CERN}
}

@article{kutsaev2021advanced,
  title={Advanced technologies for applied particle accelerators and examples of their use},
  author={Kutsaev, SV},
  journal={Technical Physics},
  volume={66},
  pages={161--195},
  year={2021},
  publisher={Springer}
}

@article{hassanein2006effects,
  title={Effects of surface damage on rf cavity operation},
  author={Hassanein, A and Insepov, Z and Norem, J and Moretti, A and Qian, Z and Bross, A and Torun, Y and Rimmer, R and Li, D and Zisman, M and others},
  journal={Physical Review Special Topics-Accelerators and Beams},
  volume={9},
  number={6},
  pages={062001},
  year={2006},
  publisher={APS}
}

@article{peralta2013demonstration,
  title={Demonstration of electron acceleration in a laser-driven dielectric microstructure},
  author={Peralta, EA and Soong, K and England, RJ and Colby, ER and Wu, Z and Montazeri, B and McGuinness, C and McNeur, J and Leedle, KJ and Walz, D and others},
  journal={Nature},
  volume={503},
  number={7474},
  year={2013},
  publisher={Nature Publishing Group UK London}
}

@article{nanni2015terahertz,
  title={Terahertz-driven linear electron acceleration},
  author={Nanni, Emilio A and Huang, Wenqian R and Hong, Kyung-Han and Ravi, Koustuban and Fallahi, Arya and Moriena, Gustavo and Dwayne Miller, RJ and K{\"a}rtner, Franz X},
  journal={Nature communications},
  volume={6},
  number={1},
  pages={8486},
  year={2015},
  publisher={Nature Publishing Group UK London}
}

@article{turnar2022waveguide,
  title={Waveguide structure based electron acceleration using terahertz pulses},
  author={Turn{\'a}r, Szabolcs and Krizs{\'a}n, Gerg{\H{o}} and Hebling, J{\'a}nos and Tibai, Zolt{\'a}n},
  journal={Optics Express},
  volume={30},
  number={15},
  pages={27602--27608},
  year={2022},
  publisher={Optica Publishing Group}
}

@article{wong2013compact,
  title={Compact electron acceleration and bunch compression in THz waveguides},
  author={Wong, Liang Jie and Fallahi, Arya and K{\"a}rtner, Franz X},
  journal={Optics express},
  volume={21},
  number={8},
  pages={9792--9806},
  year={2013},
  publisher={Optica Publishing Group}
}

@article{england2014dielectric,
  title={Dielectric laser accelerators},
  author={England, R Joel and Noble, Robert J and Bane, Karl and Dowell, David H and Ng, Cho-Kuen and Spencer, James E and Tantawi, Sami and Wu, Ziran and Byer, Robert L and Peralta, Edgar and others},
  journal={Reviews of Modern Physics},
  volume={86},
  number={4},
  pages={1337},
  year={2014},
  publisher={APS}
}

@article{liu2021thz,
  title={THz-driven dielectric particle accelerator on chip},
  author={Liu, Weihao and Sun, Li and Yu, Zijia and Liu, Yucheng and Jia, Qika and Sun, Baogen and Xu, Hongliang},
  journal={Optics Letters},
  volume={46},
  number={17},
  pages={4398--4401},
  year={2021},
  publisher={Optica Publishing Group}
}

@article{wei2018investigations,
  title={Investigations into dual-grating THz-driven accelerators},
  author={Wei, Y and Ischebeck, R and Dehler, M and Ferrari, E and Hiller, N and Jamison, S and Xia, Guoxing and Hanahoe, Kieran and Li, Yangmei and Smith, JDA and others},
  journal={Nuclear Instruments and Methods in Physics Research Section A: Accelerators, Spectrometers, Detectors and Associated Equipment},
  volume={877},
  pages={173--177},
  year={2018},
  publisher={Elsevier}
}

@article{zhang2020cascaded,
  title={Cascaded multicycle terahertz-driven ultrafast electron acceleration and manipulation},
  author={Zhang, Dongfang and Fakhari, Moein and Cankaya, Huseyin and Calendron, Anne-Laure and Matlis, Nicholas H and K{\"a}rtner, Franz X},
  journal={Physical Review X},
  volume={10},
  number={1},
  pages={011067},
  year={2020},
  publisher={APS}
}

@article{walsh2017demonstration,
  title={Demonstration of sub-luminal propagation of single-cycle terahertz pulses for particle acceleration},
  author={Walsh, DA and Lake, DS and Snedden, EW and Cliffe, MJ and Graham, DM and Jamison, SP},
  journal={Nature communications},
  volume={8},
  number={1},
  pages={421},
  year={2017},
  publisher={Nature Publishing Group UK London}
}

@article{zhao2020femtosecond,
  title={Femtosecond relativistic electron beam with reduced timing jitter from THz driven beam compression},
  author={Zhao, Lingrong and Tang, Heng and Lu, Chao and Jiang, Tao and Zhu, Pengfei and Hu, Long and Song, Wei and Wang, Huida and Qiu, Jiaqi and Jing, Chunguang and others},
  journal={Physical review letters},
  volume={124},
  number={5},
  pages={054802},
  year={2020},
  publisher={APS}
}

@article{curry2018meter,
  title={Meter-scale terahertz-driven acceleration of a relativistic beam},
  author={Curry, Emma and Fabbri, Siara and Maxson, Jared and Musumeci, Pietro and Gover, Avraham},
  journal={Physical review letters},
  volume={120},
  number={9},
  pages={094801},
  year={2018},
  publisher={APS}
}

@article{zhang2018segmented,
  title={Segmented terahertz electron accelerator and manipulator (STEAM)},
  author={Zhang, Dongfang and Fallahi, Arya and Hemmer, Michael and Wu, Xiaojun and Fakhari, Moein and Hua, Yi and Cankaya, Huseyin and Calendron, Anne-Laure and Zapata, Luis E and Matlis, Nicholas H and others},
  journal={Nature photonics},
  volume={12},
  number={6},
  pages={336--342},
  year={2018},
  publisher={Nature Publishing Group UK London}
}

@article{fallahi2016short,
  title={Short electron bunch generation using single-cycle ultrafast electron guns},
  author={Fallahi, Arya and Fakhari, Moein and Yahaghi, Alireza and Arrieta, Miguel and K{\"a}rtner, Franz X},
  journal={Physical Review Accelerators and Beams},
  volume={19},
  number={8},
  pages={081302},
  year={2016},
  publisher={APS}
}

@article{apsimon2021six,
  title={Six-dimensional phase space preservation in a terahertz-driven multistage dielectric-lined rectangular waveguide accelerator},
  author={Apsimon, {\"O}znur and Burt, Graeme and Appleby, Robert B and Apsimon, Robert J and Graham, Darren M and Jamison, Steven P},
  journal={Physical Review Accelerators and Beams},
  volume={24},
  number={12},
  pages={121303},
  year={2021},
  publisher={APS}
}

@article{huang2016terahertz,
  title={Terahertz-driven, all-optical electron gun},
  author={Huang, W Ronny and Fallahi, Arya and Wu, Xiaojun and Cankaya, Huseyin and Calendron, Anne-Laure and Ravi, Koustuban and Zhang, Dongfang and Nanni, Emilio A and Hong, Kyung-Han and K{\"a}rtner, Franz X},
  journal={Optica},
  volume={3},
  number={11},
  pages={1209--1212},
  year={2016},
  publisher={Optica Publishing Group}
}

@article{fakhari2017thz,
  title={THz cavities and injectors for compact electron acceleration using laser-driven THz sources},
  author={Fakhari, Moein and Fallahi, Arya and K{\"a}rtner, Franz X},
  journal={Physical Review Accelerators and Beams},
  volume={20},
  number={4},
  pages={041302},
  year={2017},
  publisher={APS}
}

@article{xiriai2019numerical,
  title={Numerical investigations into a THz-driven dielectric accelerator with a Bragg reflector},
  author={Xiriai, M and Shen, Baifei and Zhang, P and Aimidula, A},
  journal={Nuclear Instruments and Methods in Physics Research Section A: Accelerators, Spectrometers, Detectors and Associated Equipment},
  volume={942},
  pages={162362},
  year={2019},
  publisher={Elsevier}
}

@article{studio2010cst,
  title={CST Studio suite},
  author={Studio, CST Microwave},
  journal={Computer Simulation},
  year={2010}
}

@article{wei2017beam,
  title={Beam quality study for a grating-based dielectric laser-driven accelerator},
  author={Wei, Y and Jamison, S and Xia, Guoxing and Hanahoe, Kieran and Li, Y and Smith, JDA and Welsch, CP},
  journal={Physics of Plasmas},
  volume={24},
  number={2},
  year={2017},
  publisher={AIP Publishing}
}

@article{dai2004terahertz,
  title={Terahertz time-domain spectroscopy characterization of the far-infrared absorption and index of refraction of high-resistivity, float-zone silicon},
  author={Dai, Jianming and Zhang, Jiangquan and Zhang, Weili and Grischkowsky, Daniel},
  journal={JOSA B},
  volume={21},
  number={7},
  pages={1379--1386},
  year={2004},
  publisher={Optica Publishing Group}
}

@book{peralta2015accelerator,
  title={Accelerator on a chip: Design, fabrication, and demonstration of grating-based dielectric microstructures for laser-driven acceleration of electrons},
  author={Peralta, Edgar A},
  year={2015},
  publisher={Stanford University}
}

@article{iwaszczuk2012terahertz,
  title={Terahertz field enhancement to the MV/cm regime in a tapered parallel plate waveguide},
  author={Iwaszczuk, Krzysztof and Andryieuski, Andrei and Lavrinenko, Andrei and Zhang, X-C and Jepsen, Peter Uhd},
  journal={Optics express},
  volume={20},
  number={8},
  pages={8344--8355},
  year={2012},
  publisher={Optica Publishing Group}
}

@article{kim2010improvement,
  title={Improvement of THz coupling using a tapered parallel-plate waveguide},
  author={Kim, Sang-Hoon and Lee, Eui Su and Ji, Young Bin and Jeon, Tae-In},
  journal={Optics Express},
  volume={18},
  number={2},
  pages={1289--1295},
  year={2010},
  publisher={Optica Publishing Group}
}

@article{gallot2000terahertz,
  title={Terahertz waveguides},
  author={Gallot, Guilhem and Jamison, SP and McGowan, RW and Grischkowsky, D},
  journal={JOSA B},
  volume={17},
  number={5},
  pages={851--863},
  year={2000},
  publisher={Optica Publishing Group}
}

@article{yousefi2018silicon,
  title={Silicon dual pillar structure with a distributed Bragg reflector for dielectric laser accelerators: Design and fabrication},
  author={Yousefi, Peyman and McNeur, Joshua and Koz{\'a}k, Martin and Niedermayer, Uwe and Gannott, Florentina and Lohse, Olga and Boine-Frankenheim, Oliver and Hommelhoff, Peter},
  journal={Nuclear Instruments and Methods in Physics Research Section A: Accelerators, Spectrometers, Detectors and Associated Equipment},
  volume={909},
  pages={221--223},
  year={2018},
  publisher={Elsevier}
}

@misc{snedden2023specificationdesignenergybeam,
      title={Specification and design for Full Energy Beam Exploitation of the Compact Linear Accelerator for Research and Applications}, 
      author={E. W. Snedden and D. Angal-Kalinin and A. R. Bainbridge and A. D. Brynes and S. R. Buckley and D. J. Dunning and J. R. Henderson and J. K. Jones and K. J. Middleman and T. J. Overton and T. H. Pacey and A. E. Pollard and Y. M. Saveliev and B. J. A. Shepherd and P. H. Williams and M. I. Colling and B. D. Fell and G. Marshall},
      year={2023},
      eprint={2309.13125},
      archivePrefix={arXiv},
      primaryClass={physics.acc-ph},
      url={https://arxiv.org/abs/2309.13125}, 
}

@article{PhysRevAccelBeams.23.044801,
  title = {Design, specifications, and first beam measurements of the compact linear accelerator for research and applications front end},
  author = {Angal-Kalinin, D. and Bainbridge, A. and Brynes, A. D. and Buckley, R. K. and Buckley, S. R. and Burt, G. C. and Cash, R. J. and Castaneda Cortes, H. M. and Christie, D. and Clarke, J. A. and Clarke, R. and Cowie, L. S. and Corlett, P. A. and Cox, G. and Dumbell, K. D. and Dunning, D. J. and Fell, B. D. and Gleave, K. and Goudket, P. and Goulden, A. R. and Griffiths, S. A. and Hancock, M. D. and Hannah, A. and Hartnett, T. and Heath, P. W. and Henderson, J. R. and Hill, C. and Hindley, P. and Hodgkinson, C. and Hornickel, P. and Jackson, F. and Jones, J. K. and Jones, T. J. and Joshi, N. and King, M. and Kinder, S. H. and Knowles, N. J. and Kockelbergh, H. and Marinov, K. and Mathisen, S. L. and McKenzie, J. W. and Middleman, K. J. and Militsyn, B. L. and Moss, A. and Muratori, B. D. and Noakes, T. C. Q. and Okell, W. and Oates, A. and Pacey, T. H. and Paramanov, V. V. and Roper, M. D. and Saveliev, Y. and Scott, D. J. and Shepherd, B. J. A. and Smith, R. J. and Smith, W. and Snedden, E. W. and Thompson, N. R. and Tollervey, C. and Valizadeh, R. and Vick, A. and Walsh, D. A. and Weston, T. and Wheelhouse, A. E. and Williams, P. H. and Wilson, J. T. G. and Wolski, A.},
  journal = {Phys. Rev. Accel. Beams},
  volume = {23},
  issue = {4},
  pages = {044801},
  numpages = {37},
  year = {2020},
  month = {Apr},
  publisher = {American Physical Society},
  doi = {10.1103/PhysRevAccelBeams.23.044801},
  url = {https://link.aps.org/doi/10.1103/PhysRevAccelBeams.23.044801}
}

@article{lemery2020highly,
  title={Highly scalable multicycle THz production with a homemade periodically poled macrocrystal},
  author={Lemery, Fran{\c{c}}ois and Vinatier, Thomas and Mayet, Frank and A{\ss}mann, Ralph and Baynard, Elsa and Demailly, Julien and Dorda, Ulrich and Lucas, Bruno and Pandey, Alok-Kumar and Pittman, Moana},
  journal={Communications Physics},
  volume={3},
  number={1},
  pages={150},
  year={2020},
  publisher={Nature Publishing Group UK London}
}

@article{mosley2023large,
  title={Large-area periodically-poled lithium niobate wafer stacks optimized for high-energy narrowband terahertz generation},
  author={Mosley, Connor DW and Lake, Daniel S and Graham, Darren M and Jamison, Steven P and Appleby, Robert B and Burt, Graeme and Hibberd, Morgan T},
  journal={Optics Express},
  volume={31},
  number={3},
  pages={4041--4054},
  year={2023},
  publisher={Optica Publishing Group}
}

@article{dalton2024cryogenically,
  title={Cryogenically cooled periodically poled lithium niobate wafer stacks for multi-cycle terahertz pulses},
  author={Dalton, PJ and Shaw, CT and Bradbury, JT and Mosley, CDW and Sharma, A and Gupta, V and Bohus, J and Gupta, A and Son, J-G and F{\"u}l{\"o}p, JA and others},
  journal={Applied Physics Letters},
  volume={125},
  number={14},
  year={2024},
  publisher={AIP Publishing}
}

@article{ahr2017narrowband,
  title={Narrowband terahertz generation with chirped-and-delayed laser pulses in periodically poled lithium niobate},
  author={Ahr, Frederike and Jolly, Spencer W and Matlis, Nicholas H and Carbajo, Sergio and Kroh, Tobias and Ravi, Koustuban and Schimpf, Damian N and Schulte, Jan and Ishizuki, Hideki and Taira, Takunori and others},
  journal={Optics letters},
  volume={42},
  number={11},
  pages={2118--2121},
  year={2017},
  publisher={Optica Publishing Group}
}

@inproceedings{rentschler2024parameter,
  title={Parameter dependencies in multicycle THz generation with tunable high-energy pulse trains in large-aperture crystals},
  author={Rentschler, Christian and Matlis, NH and Demirbas, U and Zhang, Z and Pergament, M and Zukauskas, Andrius and Canalias, Carlota and Ishizuki, H and Pasiskevicius, Valdas and Laurell, Fredrik and others},
  booktitle={Nonlinear Frequency Generation and Conversion: Materials and Devices XXIII},
  volume={12869},
  pages={137--146},
  year={2024},
  organization={SPIE}
}

@inproceedings{smirnova2004photonic,
  title={Photonic band gap structures for accelerator applications},
  author={Smirnova, EI},
  booktitle={AIP Conference Proceedings},
  volume={737},
  pages={309--319},
  year={2004},
  organization={American Institute of Physics}
}

@article{lin2001photonic,
  title={Photonic band gap fiber accelerator},
  author={Lin, Xintian Eddie},
  journal={Physical Review Special Topics-Accelerators and Beams},
  volume={4},
  number={5},
  pages={051301},
  year={2001},
  publisher={APS}
}

@inproceedings{locatelli2017photonic,
  title={Photonic crystal waveguides for particle acceleration},
  author={Locatelli, A and Sorbello, G and Torrisi, G and Celona, L and De Angelis, C},
  booktitle={2017 Progress In Electromagnetics Research Symposium-Spring (PIERS)},
  pages={1008--1013},
  year={2017},
  organization={IEEE}
}

@article{mizhari2004optical,
  title = {Optical Bragg accelerators},
  author = {Mizrahi, Amit and Sch\"achter, Levi},
  journal = {Phys. Rev. E},
  volume = {70},
  issue = {1},
  pages = {016505},
  numpages = {21},
  year = {2004},
  month = {Jul},
  publisher = {American Physical Society},
  doi = {10.1103/PhysRevE.70.016505},
  url = {https://link.aps.org/doi/10.1103/PhysRevE.70.016505}
}

@article{plettner2006proposed,
  title = {Proposed few-optical cycle laser-driven particle accelerator structure},
  author = {Plettner, T. and Lu, P. P. and Byer, R. L.},
  journal = {Phys. Rev. ST Accel. Beams},
  volume = {9},
  issue = {11},
  pages = {111301},
  numpages = {12},
  year = {2006},
  month = {Nov},
  publisher = {American Physical Society},
  doi = {10.1103/PhysRevSTAB.9.111301},
  url = {https://link.aps.org/doi/10.1103/PhysRevSTAB.9.111301}
}

@article{liu2020microscale,
  title={Microscale laser-driven particle accelerator using the inverse Cherenkov effect},
  author={Liu, Weihao and Yu, Zijia and Sun, Li and Liu, Yucheng and Jia, Qika and Xu, Hongliang and Sun, Baogen},
  journal={Physical Review Applied},
  volume={14},
  number={1},
  pages={014018},
  year={2020},
  publisher={APS}
}

@article{palfalvi2014evanescent,
  title={Evanescent-wave proton postaccelerator driven by intense THz pulse},
  author={P{\'a}lfalvi, L{\'a}szl{\'o} and F{\"u}l{\"o}p, J{\'o}zsef Andr{\'a}s and T{\'o}th, Gy and Hebling, J{\'a}nos},
  journal={Physical Review Special Topics-Accelerators and Beams},
  volume={17},
  number={3},
  pages={031301},
  year={2014},
  publisher={APS}
}

@article{leedle2015dielectric,
  title={Dielectric laser acceleration of sub-100 keV electrons with silicon dual-pillar grating structures},
  author={Leedle, Kenneth J and Ceballos, Andrew and Deng, Huiyang and Solgaard, Olav and Fabian Pease, R and Byer, Robert L and Harris, James S},
  journal={Optics letters},
  volume={40},
  number={18},
  pages={4344--4347},
  year={2015},
  publisher={Optical Society of America}
}

@article{breuer2014dielectric,
  title={Dielectric laser acceleration of electrons in the vicinity of single and double grating structures—theory and simulations},
  author={Breuer, John and McNeur, Joshua and Hommelhoff, Peter},
  journal={Journal of Physics B: Atomic, Molecular and Optical Physics},
  volume={47},
  number={23},
  pages={234004},
  year={2014},
  publisher={IOP Publishing}
}

@article{niedermayer2018alternating,
  title={Alternating-phase focusing for dielectric-laser acceleration},
  author={Niedermayer, Uwe and Egenolf, Thilo and Boine-Frankenheim, Oliver and Hommelhoff, Peter},
  journal={Physical review letters},
  volume={121},
  number={21},
  pages={214801},
  year={2018},
  publisher={APS}
}

@article{chlouba2023coherent,
  title={Coherent nanophotonic electron accelerator},
  author={Chlouba, Tom{\'a}{\v{s}} and Shiloh, Roy and Kraus, Stefanie and Br{\'u}ckner, Leon and Litzel, Julian and Hommelhoff, Peter},
  journal={Nature},
  volume={622},
  number={7983},
  pages={476--480},
  year={2023},
  publisher={Nature Publishing Group UK London}
}

@article{xiriai2024numerical,
  title={Numerical investigation of a THz-driven dielectric accelerator with a Bragg-reflector for accelerating sub-relativistic electron beams},
  author={Xiriai, M and Aimidula, Aimierding and Bake, Mamat Ali and Zhang, Ping},
  journal={Physica Scripta},
  year={2024}
}

@article{thompson2008breakdown,
  title={Breakdown Limits on Gigavolt-per-Meter Electron-Beam-Driven Wakefields in Dielectric Structures},
  author={Thompson, MC and Badakov, H and Cook, AM and Rosenzweig, JB and Tikhoplav, R and Travish, G and Blumenfeld, I and Hogan, MJ and Ischebeck, R and Kirby, N and others},
  journal={Physical review letters},
  volume={100},
  number={21},
  pages={214801},
  year={2008},
  publisher={APS}
}

@article{chefonov2017giant,
  title={Giant self-induced transparency of intense few-cycle terahertz pulses in n-doped silicon},
  author={Chefonov, OV and Ovchinnikov, AV and Romashevskiy, SA and Chai, X and Ozaki, T and Savel’ev, AB and Agranat, MB and Fortov, VE},
  journal={Optics Letters},
  volume={42},
  number={23},
  pages={4889--4892},
  year={2017},
  publisher={Optical Society of America}
}

@article{o2019conductivity,
  title={Conductivity induced by high-field terahertz waves in dielectric material},
  author={O’Shea, Brendan D and Andonian, Gerard and Barber, Samuel K and Clarke, Christine Isabel and Hoang, Phuc D and Hogan, Mark J and Naranjo, Brian and Williams, Oliver Brian and Yakimenko, Vitaly and Rosenzweig, James B},
  journal={Physical review letters},
  volume={123},
  number={13},
  pages={134801},
  year={2019},
  publisher={APS}
}

@article{tarekegne2017impact,
  title={Impact ionization dynamics in silicon by MV/cm THz fields},
  author={Tarekegne, Abebe T and Hirori, Hideki and Tanaka, Koichiro and Iwaszczuk, Krzysztof and Jepsen, Peter U},
  journal={New Journal of Physics},
  volume={19},
  number={12},
  pages={123018},
  year={2017},
  publisher={IOP Publishing}
}

@inproceedings{zhu2016robustness,
  title={Robustness of various metals against high THz field induced damage},
  author={Zhu, Jianfei and Iwaszczuk, Krzysztof and Tarekegne, Abebe T and Ma, Yungui and Jepsen, Peter U},
  booktitle={2016 41st International Conference on Infrared, Millimeter, and Terahertz waves (IRMMW-THz)},
  pages={1--2},
  year={2016},
  organization={IEEE}
}

@article{v2018damage,
  title={Damage threshold of Ni thin film by terahertz pulses},
  author={V. Chefonov, O and V. Ovchinnikov, A and A. Evlashin, S and B. Agranat, M},
  journal={Journal of Infrared, Millimeter, and Terahertz Waves},
  volume={39},
  number={11},
  pages={1047--1054},
  year={2018},
  publisher={Springer}
}

@inproceedings{chefonov2018interaction,
  title={Interaction of High-Power Terahertz Radiation with Metallic Films},
  author={Chefonov, OV and Ovchinnikov, AV and Ashitkov, SI and Evlashin, SA and Kondratenko, PS and Agranat, MB and Fortov, VE},
  booktitle={EPJ Web of Conferences},
  volume={195},
  pages={07001},
  year={2018},
  organization={EDP Sciences}
}

@article{agranat2018damage,
  title={Damage in a thin metal film by high-power terahertz radiation},
  author={Agranat, MB and Chefonov, OV and Ovchinnikov, AV and Ashitkov, SI and Fortov, VE and Kondratenko, PS},
  journal={Physical Review Letters},
  volume={120},
  number={8},
  pages={085704},
  year={2018},
  publisher={APS}
}

@article{atakaramians2013terahertz,
  title={Terahertz dielectric waveguides},
  author={Atakaramians, Shaghik and Afshar V, Shahraam and Monro, Tanya M and Abbott, Derek},
  journal={Advances in Optics and Photonics},
  volume={5},
  number={2},
  pages={169--215},
  year={2013},
  publisher={Optical Society of America}
}

@article{andrews2014microstructured,
  title={Microstructured terahertz waveguides},
  author={Andrews, Steven R},
  journal={Journal of Physics D: Applied Physics},
  volume={47},
  number={37},
  pages={374004},
  year={2014},
  publisher={IOP Publishing}
}

@article{wu2025review,
  title={Review on the terahertz transmission devices and their applications: From metal waveguides to terahertz fibers},
  author={Wu, Ye-Qing and Chen, Ming-Yang and Dai, Zi-Jie},
  journal={Optics \& Laser Technology},
  volume={183},
  pages={112339},
  year={2025},
  publisher={Elsevier}
}

@article{yu2023megaelectronvolt,
  title={Megaelectronvolt electron acceleration driven by terahertz surface waves},
  author={Yu, Xie-Qiu and Zeng, Yu-Shan and Song, Li-Wei and Kong, De-Yin and Hao, Si-Bo and Gui, Jia-Yan and Yang, Xiao-Jun and Xu, Yi and Wu, Xiao-Jun and Leng, Yu-Xin and others},
  journal={Nature Photonics},
  volume={17},
  number={11},
  pages={957--963},
  year={2023},
  publisher={Nature Publishing Group UK London}
}

@article{ying2024high,
  title={High gradient terahertz-driven ultrafast photogun},
  author={Ying, Jianwei and He, Xie and Su, Dace and Zheng, Lingbin and Kroh, Tobias and Rohwer, Timm and Fakhari, Moein and Kassier, G{\"u}nther H and Ma, Jingui and Yuan, Peng and others},
  journal={Nature Photonics},
  volume={18},
  number={7},
  pages={758--765},
  year={2024},
  publisher={Nature Publishing Group UK London}
}

@article{nix2024terahertz,
  title={Terahertz-driven acceleration of subrelativistic electron beams using tapered rectangular dielectric-lined waveguides},
  author={Nix, Laurence JR and Bradbury, Joseph T and Shaw, Christopher T and Hibberd, Morgan T and Graham, Darren M and Appleby, Robert B and Burt, Graeme and Letizia, Rosa and Jamison, Steven P},
  journal={Physical Review Accelerators and Beams},
  volume={27},
  number={4},
  pages={041302},
  year={2024},
  publisher={APS}
}

@article{mbonye2012study,
  title={Study of the impedance mismatch at the output end of a THz parallel-plate waveguide},
  author={Mbonye, Marx and Mendis, Rajind and Mittleman, Daniel M},
  journal={Applied Physics Letters},
  volume={100},
  number={11},
  year={2012},
  publisher={AIP Publishing}
}

@article{othman2019parallel,
  title={Parallel-plate waveguides for terahertz-driven MeV electron bunch compression},
  author={Othman, Mohamed AK and Hoffmann, Matthias C and Kozina, Michael E and Wang, XJ and Li, Renkai K and Nanni, Emilio A},
  journal={Optics express},
  volume={27},
  number={17},
  pages={23791--23800},
  year={2019},
  publisher={Optical Society of America}
}

@article{mueckstein2013mode,
  title={Mode interference and radiation leakage in a tapered parallel plate waveguide for terahertz waves},
  author={Mueckstein, R and Navarro-Cia, Miguel and Mitrofanov, O},
  journal={Applied Physics Letters},
  volume={102},
  number={14},
  year={2013},
  publisher={AIP Publishing}
}

\end{document}